\documentclass[aps,twocolumn,notitlepage,superscriptaddress]{revtex4-1}

\usepackage{bm}
\usepackage{graphicx}
\usepackage{latexsym}
\usepackage{revsymb}
\usepackage{amsmath,amssymb}
\usepackage{mathtools}
\usepackage{mathrsfs}
\usepackage[colorlinks=true,citecolor=blue,linkcolor=blue]{hyperref}
\usepackage{bbding}
\usepackage[caption=false]{subfig} 
\usepackage{float}
\usepackage{subfloat}
\usepackage{graphicx}
\usepackage[section]{placeins}
\usepackage{csquotes}
\usepackage{relsize}
\usepackage{enumitem}
\usepackage[Symbol]{upgreek}
\DeclareGraphicsExtensions{.pdf,.png,.jpg}

\LetLtxMacro{\OldSqrt}{\sqrt}
\newcommand{\ClosedSqrt}[1][\hphantom{3}]{\def\DHLindex{#1}\mathpalette\DHLhksqrt}
\makeatletter
    \newcommand*\bold@name{bold}
    \def\DHLhksqrt#1#2{%
        \setbox0=\hbox{$#1\OldSqrt{#2\,}$}\dimen0=\ht0\relax%
        \advance\dimen0-0.2\ht0\relax
        \setbox2=\hbox{\vrule height\ht0 depth -\dimen0}%
        {\hbox{$#1\expandafter\OldSqrt\expandafter[\DHLindex]{#2\,}$}
        \lower\ifx\math@version\bold@name0.6pt\else0.4pt\fi\box2}
    }
    \renewcommand*{\sqrt}[2][\ ]{\ClosedSqrt[\leftroot{-2}\uproot{1}#1]{#2}\kern0.1em} 
\makeatother

\renewcommand\vec{\mathbf}

\begin{document}

\title{Extreme plasmons}

\author{Aakash A. Sahai}
\affiliation{Department of Electrical Engineering, University of Colorado Denver, Denver, CO 80204}
\email[corresponding author: ~]{aakash.sahai@ucdenver.edu}

\begin{abstract}
Nanosciences largely rely on plasmons which are quasiparticles constituted by collective oscillations of quantum electron gas composed of conduction band electrons that occupy discrete quantum states. Our work has introduced non-perturbative plasmons with oscillation amplitudes that approach the extreme limit set by breakdown in characteristic coherence. In contrast, conventional plasmons are small-amplitude oscillations. Controlled excitation of extreme plasmons modeled in our work unleashes unprecedented Petavolts per meter fields. In this work, an analytical model of this new class of plasmons is developed based on quantum kinetic framework. A controllable extreme plasmon, the surface ``crunch-in'' plasmon, is modeled here using a modified independent electron approximation which takes into account the quantum oscillation frequency. Key characteristics of such realizable extreme plasmons that unlock unparalleled possibilities, are obtained.
\end{abstract}


\maketitle

\section{Introduction}

Plasmons \cite{Pines-PhysRev-1953, Pines-Bohm-PhysRev-1952, Ritchie-surface-plasmon} underpin nanosciences \cite{plasmonics-1} by making possible nanometric energy confinement. Nanometric size of these quasiparticles constituted by collective oscillations of delocalized, conduction band electron gas is realizable only due to the innately quantum nature of the gas which allows ultra-high densities, $n_0\simeq10^{24}\mathrm{cm^{-3}}$ that enforce a characteristic wavelength:
\begin{align}\label{eq:plasmonic-wavelength}
\begin{split} 
\lambda_{Q}\lesssim\frac{30}{\sqrt{n_0\mathrm{(10^{24}cm^{-3}})}} ~ \mathrm{nm}.
\end{split}
\end{align}
The very existence of the quantum electron gas (densities unattainable for a classical gas) and thereby nanoscale plasmons is dependent on quantum effects. These nonclassical properties are essential for collisionless dynamics of conduction electrons, as determined by Bloch's theorem \cite{Bloch-1933} (electron-lattice interactions) and Pauli's exclusion principle \cite{Pauli-spin-statistics} (electron-electron). 

The conduction electron gas maintains its quantum character with electron states valid only at discrete energies, $\mathcal{E}_\vec{k}$ and wavevectors, $\vec{k_{\ell}}$ (unlike a classical gas) as long as the ionic lattice exists. These discrete states are occupied in accordance with Fermi-Dirac statistics which makes the quantum electron gas free of collisions with an ideal, stationary lattice. The nonclassical distribution along with nearly non-interacting fermions gives rise to a Fermi electron gas, a quantum entity which is a prerequisite for plasmons.

\begin{figure}[!htb]
\centering
   \includegraphics[width=\columnwidth]{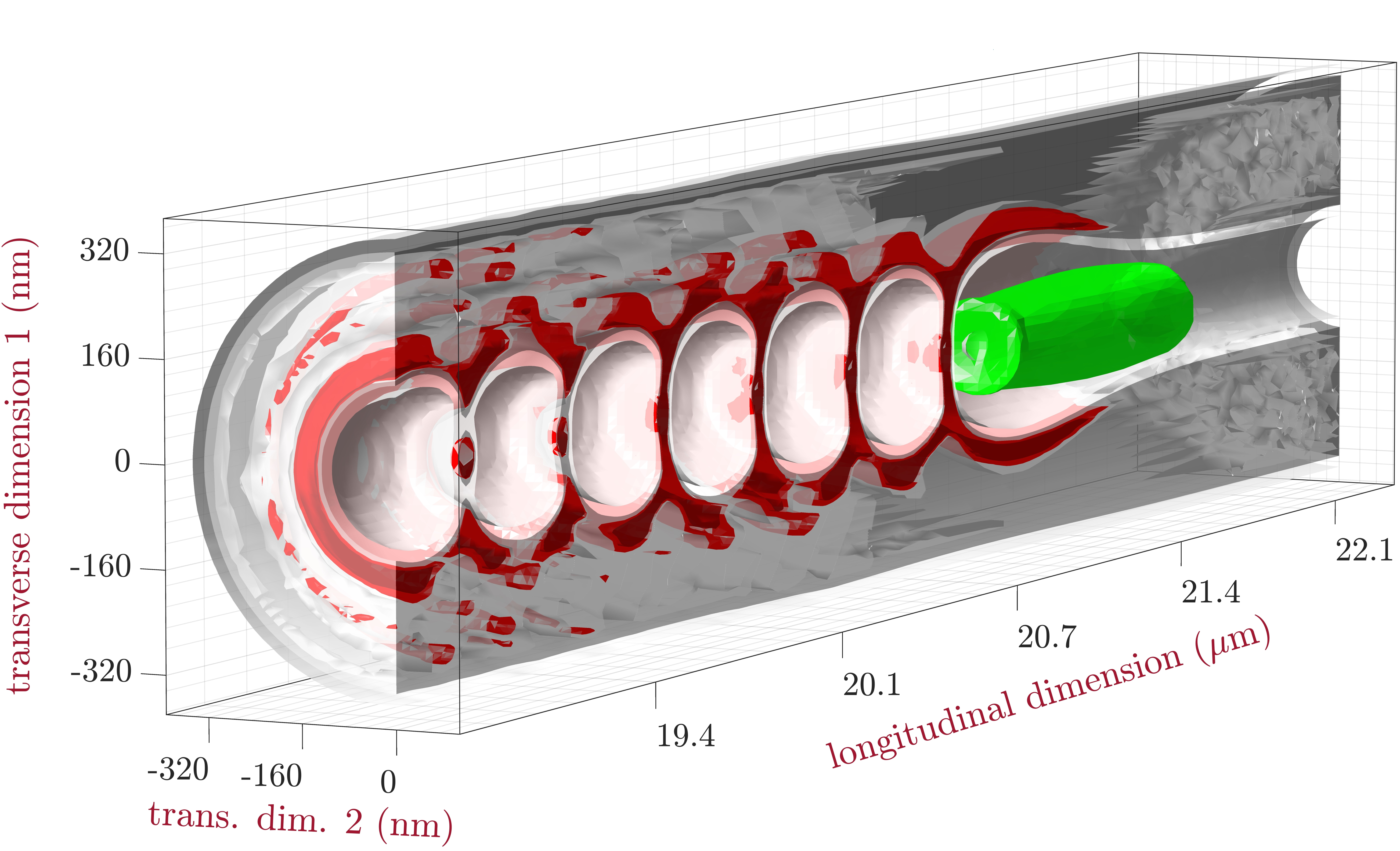}
   \caption{Extreme plasmon excited in a conductive tube by an electron bunch (green envelope) launched inside it,  illustrated in 3D using particle-tracking simulation. The conduction electrons in a tube with core radius: $r_t\rm=100nm$, wall thickness: $\Delta w\rm=250nm$ and electron density: $n_0=\rm 2\times10^{22}cm^{-3}$ are relativistically excited. The red (denser) and gray surfaces show electron density variations. Electron trajectories approach $\lambda_{Q}\simeq 200{\rm nm}$ and tunnel across the tube surface to ``crunch-in''. The bunch that has here propagated about $\rm20\mu m$ inside the tube is initially Gaussian with: $\sigma_z\rm=400nm$, $\sigma_r\rm=250nm$ and density, $n_{b0}=\rm 5\times10^{21}cm^{-3}$.}
\label{fig:3D-crunch-in-plasmon}
\end{figure}

Our work has uncovered a new class of plasmons in the large-amplitude limit \cite{plasmonics-ieee-2021,spie-2021,pct-2021,plasmonics-ieee-2022,nanofocusing-2022, ion-wake, crunch-in-regime, jinst-2023}. Large-amplitude, non-perturbative plasmons have a trajectory amplitude ($\delta$) of collective oscillations that approaches the wavelength, $\delta\simeq\lambda_{Q}$. The extreme limit of oscillation amplitude is set by breakdown in mutual phase coherence (onset of collisions) which is critical for the quintessential ``collective'' nature of plasmons. In contrast, conventional plasmons are sustained by oscillations of small amplitude, $\delta\ll\lambda_{Q}$. Furthermore, while  conventional plasmons are made up of excited states with electron velocities relatively close to the Fermi velocity, $v_F$; excited quantum states underlying large-amplitude plasmons correspond with relativistic electron momenta, ${\bf p}\simeq m_ec\gg p_F$. 

Extreme plasmons introduced by our work gain prominence as they open unprecedented PetaVolts per meter (PV/m) electromagnetic fields \cite{plasmonics-ieee-2021, spie-2021, nanofocusing-2022} (sec. \ref{subsec:quantum-coherence-limit}). As this unique quantum technology brings about a PV/m {\it extreme field frontier}, it promises unparalleled possibilities. A few of which include nano-wiggler based gamma-ray lasers \cite{plasmonics-ieee-2021, ginzburg-2003, pct-2021} and opening the vacuum (field-driven vacuum polarization) \cite{jinst-2023, schwinger-limit}.

Our efforts \cite{slac-2023,slac-2022, slac-2020} are aimed at controlled excitation of extreme plasmons. Controlled nanometric confinement of electromagnetic energy is only possible by preserving the ingrained quantum nature critical to sustain collisionless oscillations. This quantum nature vital for plasmons is maintained by the ionic lattice which is preserved in our model by exciting plasmons using high-density charged particle beams \cite{plasmonics-ieee-2021, pct-2021}. Due to bunching characteristics, particle accelerators intrinsically rid the beams of any significant pre-pulses capable of disrupting the lattice (contrary to high-intensity lasers).

Further control over the interaction is enabled by a specific type of extreme plasmon, which based upon the key requirements described in sec.\ref{sec:controlled-excitation} has been identified and introduced in our work as the surface crunch-in plasmon. However, unlike perturbative surface plasmons \cite{Ritchie-surface-plasmon}, plasmons in the large-amplitude limit have highly localized, small spatial scale collective electron accumulation (large collective wavevector, $\vec{k}\simeq 2\pi\lambda_Q^{-1}$) excited by relativistic  momenta ($\bf p$). These characteristics invalidate customary fluid model based on initialized constitutive parameters and necessitate kinetic modeling supported by particle-tracking simulations (in Fig.\ref{fig:3D-crunch-in-plasmon}). In Fig.\ref{fig:3D-crunch-in-plasmon}, where $\lambda_Q$ is around 200 nm; conduction electrons are collectively displaced by $\delta \simeq 100 {\rm nm}$ and accumulate over a few tens of nm, at the maxima of their trajectories. 

In this work, extreme plasmons are analytically modeled from the first principles. These principles and the framework they shape are laid out in sec.\ref{sec:framework}. In particular, our quantum kinetic model is based on the {\it Wigner quantum phase-space} formalism. Using this formalism under a modified {\it independent electron approximation} which accounts for the quantum oscillation frequency, $\omega_Q$ (see sec.\ref{subsec:quantum-small-plasmons}), we model a surface ``crunch-in'' plasmon collisionlessly excited on the inner surface of conductive tubes, as depicted in Fig.\ref{fig:3D-crunch-in-plasmon}. The kinetic model developed in sec.\ref{sec:kinetic-model} is used to estimate the fields in sec.\ref{sec:field-estimation}.

The extreme plasmon model, particularly the surface crunch-in plasmon forms the basis of our ongoing experimental effort \cite{slac-2023,slac-2022, slac-2020} (initially using tunable, semiconductor plasmons \cite{plasmonics-ieee-2022}).

\section{Quantum framework}
\label{sec:framework}

Quantum systems, here the conduction electron gas, are made up of entities that occupy discrete states. These discrete levels are modified in an excited state such as a plasmon. As extreme plasmons are sustained by strong excitation of quantum electron gas, the statistical distribution of electron states significantly diverges from the initial Fermi-Dirac statistics, making linear theory inapplicable. A quantum kinetic framework is, thus, put forth here to account for this. 

It is well known that Boltzmann (or Vlasov) equation based classical kinetic models applied to quantum systems with discrete energy states predict unphysical effects conflicting with experimental observations.

\subsection{Quantum system: Degeneracy and Correlation}
\label{subsec:quantum-state}
The quantum state of a system is defined by its characteristics at equilibrium. Quantum degeneracy (parameter, $\chi$) and quantum correlation (parameter, $\Gamma$) are two such key characteristics.

{\bf Quantum degeneracy parameter:} $\chi = \frac{8}{3\sqrt{\pi}}\left(\frac{\mathcal{E}_F}{k_BT}\right)^{3/2}$ quantifies Fermi energy, $\mathcal{E}_F = \frac{\hbar^2}{2m_e}(3\pi^2 n_0)^{2/3}$ relative to thermal energy, $k_BT$ (Maxwellian distribution). The conduction electron gas with densities $10^{18-24}{\rm cm}^{-3}$ (doped semiconductors to metals) is degenerate, $\chi \gtrsim 1$ at $T=300K$.

{\bf Quantum correlation parameter:} $\Gamma=8\frac{2^{1/3}}{3\pi^2}d_0a_0^{-1}$, which compares the inter-particle spacing, $d_0=(4\pi^3)^{1/3}\left(3\pi^2 n_0\right)^{-1/3}$ against Bohr radius, $a_0$ (or the deBroglie wavelength); characterizes the conduction electron gas (after establishing degeneracy) as being strongly correlated with $\Gamma\gtrsim1$. 

The energy arising from correlation further differentiates the quantum state from a classical one. Correlation energy is defined as the difference in energy of a many-body electron state such as the conduction electron gas relative to its single-electron approximation.

While a quantum (degenerate and correlated) system becomes more strongly correlated with increase in density, $n_0$, a classical one contrarily departs from its ideal behavior. This is a uniquely quantum characteristic where the kinetic energy increases with density. 

\subsection{Wigner quantum phase-space} 
The dynamics of any many-body quantum system with N quantum entities such as the quantum electron gas in this work, is fully describable by suitably transforming (Wigner-Weyl) the quantum variables representing the ensemble over the real-space $\mathbb{R}^{3N}$. Instead of exact determination of many-body wavefunction, the density matrix, $\rho(\vec{r},\vec{r'},t)$ which allows determination of the statistical distribution of states, is used to obtain the {\em Wigner function}, $f_W$ (in Complex space, $\mathbb{C}$):
\begin{align}\label{eq:Wigner-Weyl-transform}
\begin{split} 
& f_W(\vec{r},\vec{p},t) = \int_{\mathbb{R}^{3N}\rightarrow\mathbb{C}} d^3s~~e^{-is\frac{\vec{p}}{\hbar}}~ \rho\left(\vec{r}+\frac{s}{2}, \vec{r}-\frac{s}{2}, t\right). \\
& \rho = \int_{\mathbb{C}\rightarrow\mathbb{R}^{3N}}\int_{a,b} \frac{f_W(\vec{r},\vec{p})}{(2\pi\hbar)^{3N}} \left(e^{i[a(Q-\vec{r})+b(P-\vec{p})]}\right) \text{d}x\, \text{d}p\, \text{d}a\, \text{d}b.
\end{split}
\end{align}
This representation overcomes the limitations imposed by Heisenberg's uncertainty principle. It allows treatment of quantum mechanics in phase-space by simultaneous representation of the distribution of non-commuting variables, here instantaneous position ($\hat{Q}$) and momentum ($\hat{P}$). Although, $f_W$ is a quasi probability distribution function, it allows modeling the evolution of a quantum system using phase-space formalism. Density, $n(\vec{r},t)$ and momentum, $n(\vec{p},t)$ distributions are obtained from the Wigner function, as follows:
 \begin{align}\label{eq:density-momentum-distribution}
\begin{split} 
\int d^3p ~ f_W(\vec{r}, \vec{p}, t) = n(\vec{r}, t).\\
\int d^3r ~ f_W(\vec{r}, \vec{p}, t) = n(\vec{p}, t).
\end{split}
\end{align}

The evolution of Wigner distribution function is described by the {\em Wigner equation} \cite{Wigner}. 
\begin{align}\label{eq:wigner-eq}
\begin{split} 
\frac{\partial f_W}{\partial t} + \frac{\hbar\vec{k_\ell}}{2m^*}\frac{\partial f_W}{\partial \vec{r}} = \int d\vec{k_\ell'} ~ \phi_W(\vec{r},\vec{k_\ell'}-\vec{k_\ell},t)f_W(\vec{r},\vec{k_\ell'},t)
\end{split}
\end{align}
The Wigner transport equation, Eq.\ref{eq:wigner-eq} describes ballistic carrier transport. Further correction terms are needed to account for collisional dynamics. Here, $\vec{k_\ell}$ ($\hat{p}=-i\hbar\vec{\nabla}$) is the wavevector which determines the lattice or crystal momentum of individual electrons with effective mass, $m_e^*$. Naturally, when conduction electrons that sustain the extreme plasmon individually attain large $\vec{k_\ell}$ the collective wavevector, $\vec{k}$ also becomes large.

The electromagnetic force due to long-range Coulomb interaction (due to non-equilibrium distribution in space) in the Wigner equation is accounted for by an electrostatic potential, $\phi_W$ (using mean-field approximation) which is determined by Poisson equation:
\begin{align}\label{eq:Poisson-eq}
\begin{split} 
\nabla^2\phi_W(\vec{r}, t) = 4\pi e (n(\vec{r}, t) - n_0).
\end{split}
\end{align}

For the quantum electron gas, the Wigner-Poisson system provides a framework similar to the Vlasov-Poisson system for classical gasses.

\begin{figure}[!htb]
\centering
   \includegraphics[width=0.9\columnwidth]{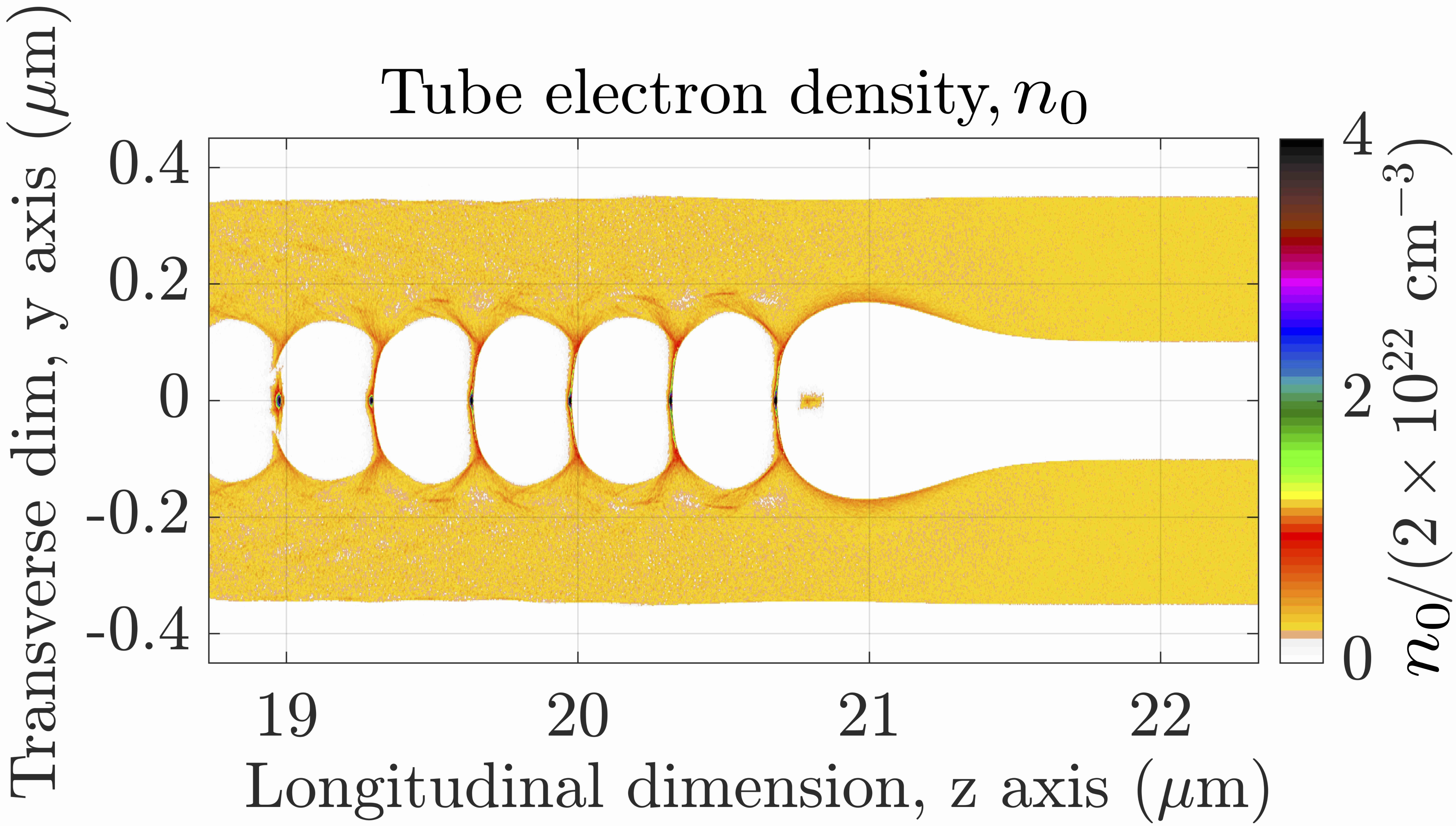}
   \caption{Cross-section of the surface crunch-in plasmon in a conductive tube from 3D particle-tracking simulation same as in Fig.\ref{fig:3D-crunch-in-plasmon} using \cite{epoch-pic,xsede-rmacc-citation}.}
\label{fig:cross-section-3D-crunchin-plasmon}
\end{figure}

In our work, the quantum electron gas on a surface is strongly excited such that the displacement amplitude of  constituent conduction electrons approaches the wavelength, in Eq.\ref{eq:plasmonic-wavelength}. The resulting large-amplitude trajectories cannot be perturbatively treated and linearized. Such trajectories underlying extreme plasmons are discernible from Fig.\ref{fig:cross-section-3D-crunchin-plasmon} which shows the planar cross-section of electron density (of the 3D simulation in Fig.\ref{fig:3D-crunch-in-plasmon}). 

Consequently, large-amplitude quantum gas oscillations cannot be modeled using fluid theory (tracking macroscopic quantities such as moments of phase-space) which uses macroscopic material properties and classical hydrodynamic model \cite{Kronig-Korringa-1943}. Similarly, small-amplitude quantum mechanical framework \cite{Pines-Bohm-PhysRev-1952} based upon perturbative expansion around smallness of $\alpha$ (see sec.\ref{subsec:quantum-small-plasmons}), is not applicable to extreme plasmons.

Our analytical model below, therefore, uses quantum kinetic theory to model large-amplitude plasmons. Particularly, we build upon the small-amplitude plasmon model \cite{Pines-Bohm-PhysRev-1952} that matches experimental observations.

\subsection{Modified independent electron approximation}
\label{quantum-kinetic-approx}
In kinetic modeling of quantum systems, under certain conditions it is sufficient to calculate the trajectory of a single electron. When the energy of electron-electron interaction (already minimized by Pauli's exclusion principle) becomes small relative to kinetic energy and electron-lattice potential terms, particularly, in the large-amplitude limit, the many-body electron model can be decomposed into individual electron dynamics with corrections for correlations.

The independent electron approximation allows for the large-amplitude surface crunch-in plasmon model based upon evolution of Wigner distribution function, $f_W(\vec{r},\vec{p},t)$ in Eq.\ref{eq:wigner-eq} to be fully accounted for by the dynamics of individual electrons. As this approximation is applicable over collisionless timescale, our model is applicable to the dynamics before collisions begin to dominate.

Accordingly, the quantum kinetic model especially at the extreme limits is here approximated by a modified single-particle kinetic equation utilizing the quantum factor, $\mathscr{F}_Q(\vec{k},\vec{p})$ introduced in sec.\ref{subsec:quantum-small-plasmons}.

Formal evaluation of the efficacy of assumptions by comparing experiments against treatment using quantum kinetic model and a modified single-particle approach adopted here, is part of our future work.

\subsection{Small-amplitude plasmons: Quantum model}
\label{subsec:quantum-small-plasmons}
Quantum mechanical framework of small-amplitude plasmons was developed by Pines et. al. \cite{Pines-Bohm-PhysRev-1952, Pines-PhysRev-1953} to explain experimental observations. As the electronic states whether in the ground (equilibrium) or the excited (plasmon) state of conduction electrons, are valid only at discrete wavevectors, such a framework is indispensable. 

The dynamics of quantum electrons, particularly their organized behavior which gives rise to collective electron oscillations, is fundamentally a many-body quantum phenomena. Analysis of this many-body quantum state is simplified in \cite{Pines-PhysRev-1953} by using a canonical transform to the basis of collective wavevector, $\vec{k}$.

This transformation allows separating the Hamiltonian into three major, nearly independent terms:
\begin{align}\label{eq:Hamiltonian-quantum}
\begin{split} 
\mathcal{H} = \mathcal{H}_{\rm part}+\mathcal{H}_{\rm coll}+\mathcal{H}_{s. r.} 
\end{split}
\end{align}
Eq.\ref{eq:Hamiltonian-quantum} explicitly recognizes the existence of plasmons using the term $\mathcal{H}_{\rm coll}$. The discrete wavevector basis is suitable for separation of terms contributing to $\mathcal{H}$, because plasmons are only sustained below a {\em critical wavenumber}, $k<k_c$. This critical wavenumber, $k_c$ is close to the {\em Thomas-Fermi screening wavenumber}:
\begin{align}\label{eq:plasmon-critical-wavenumber}
\begin{split} 
k_c^2 \simeq k_{\rm TF}^2 = k_F^2 \left( \frac{16}{3\pi^2}\right)^{2/3} \frac{r_s}{a_0}
\end{split}
\end{align}
This critical wavenumber motivates the canonical transformation as it allows separation of terms in the small-amplitude limit. Using this transformation, analysis of plasmons becomes independent of individual electron dynamics that primarily occurs over $k>k_c$, represented by $\mathcal{H}_{\rm part}$ (individual particle states) and $\mathcal{H}_{s. r.}$ (short range). Here, the Fermi wavevector, $k_F=(3\pi^2 n_0)^{1/3}$ represents electron momentum at the Fermi energy, $E_F=\hbar^2k_F^2(2m_e)^{-1}$, $r_s$ is radius of the Wigner-Seitz sphere and $a_0$ the Bohr radius.

The energy of plasmons ($\mathcal{H}_{\rm coll}$) is required to be higher than the ground-state energy (itself greater than the Fermi energy) of the electron assembly to eliminate their spontaneous excitation. So, when the quantum electron gas is externally excited to sustain plasmons, the term $\mathcal{H}_{\rm coll}$ in Eq.\ref{eq:Hamiltonian-quantum} becomes dominant. 

Experimentally verifiable quantities such as the frequency of plasmon, $\omega_{Q}$ using the quantum model in \cite{Pines-Bohm-PhysRev-1952} were obtained in the small-amplitude limit by perturbatively expanding about a small parameter, $\alpha$:
\begin{align}\label{eq:alpha-QM}
\begin{split} 
\alpha = \left\langle \left(\frac{\vec{k} \cdot \vec{p}}{m\omega}\right)^2 \right\rangle
\end{split}
\end{align}
where, the parameter $\alpha$ is averaged over the particle momenta, $\vec{p}$ and wavevectors of the collective field (or collective density profile), $\vec{k}$ with $\omega$ being the frequency of the collective oscillations. The frequency of  conduction electrons, $\omega$ was defined using quantum dispersion relation obtained based upon small $\alpha$. 
The average plasmon frequency, $\left\langle\omega_{Q}\right\rangle$ from the perturbative expansion about small $\alpha$ is:
\begin{align}\label{eq:plasmon-frequency-small-amplitude}
\begin{split} 
\left\langle\omega_{Q}\right\rangle \simeq \left( 1 + 3\alpha\left[1+ \frac{3}{10}\beta^2\right] \right) \omega_{p} \equiv  \mathscr{F}_Q(\vec{k},\vec{p}) ~ \omega_{p}
\end{split}
\end{align}
where, $\omega_{p}=\sqrt{\frac{n_0 e^2}{\epsilon_0 m_e}}$ is the classical electron (plasma) oscillation frequency, $\beta = \frac{k_c}{k_F}$, and $\alpha \simeq \frac{1}{2} ~ \beta^2 ~ \frac{a_0}{r_s}$. The plasmon frequency is, therefore, material and ($\vec{k}$,$\vec{p}$)-dependent as evident from the term $3\alpha$.

{\bf Quantum factor:} In this work, we introduce a quantum factor $\mathscr{F}_Q(\alpha,\beta,\vec{k},\vec{p})\equiv \mathscr{F}_Q(\vec{k},\vec{p})$. The characteristics of quantum electron oscillations ($\omega_{Q}$) are quite unlike the oscillations in a classical plasma ($\omega_{p}$) which has randomly distributed ions (no quantum effects) and Maxwell-Boltzmann distribution (energy continuum). 

In \cite{Pines-PhysRev-1953}, characteristics of plasmons measured in experiments were demonstrated to be in agreement with the quantum model (as opposed to a classical one).

{\bf Non-perturbative plasmons:} Extreme plasmons are not amenable to perturbative expansion because of their large oscillation amplitude. Such oscillations have large wavevectors, $\vec{k}$ as is evident from highly localized density accumulation. These large collective wavevectors are a result of relativistic momentum, $\vec{p}$.

Due to the lack of any experimental investigations of extreme plasmons, their underlying non-perturbative, quantum mechanical framework and their characteristic quantities such as oscillation frequency of extreme plasmons, $\omega_{Q}$, quantum properties are incorporated in our model using the quantum factor, $\mathscr{F}_Q({\vec{k},\vec{p}})$. $\mathscr{F}_Q$ ($>1$ even in the small-amplitude limit) is not merely a constant factor \cite{Pines-PhysRev-1953}, it depends upon material properties and scales with $\vec{k}$ and $\vec{p}$ of the excited states.

In the large-amplitude quantum limit, the plasmon frequency and coherence-limit of quantum electron oscillations are obtained by extending into the quantum domain using $\mathscr{F}_Q$. Our experiments \cite{plasmonics-ieee-2022, slac-2023, slac-2022, slac-2020} will comprehensively address these considerations.

{\bf Preceding models:} Quantum mechanical treatment of conduction electron oscillations in \cite{Pines-PhysRev-1953} was preceded by hydrodynamic theory of Kronig and Korringa \cite{Kronig-Korringa-1943} which treated the quantum electron gas as a classical fluid. This model accounted for collisions as viscous friction which led to the erroneous conclusion that collision rate increased with the velocity of oscillating electrons.

Subsequently, Kramers proposed an alternative model \cite{Kramers-1947} in line with an earlier classical kinetic treatment by Drude which was developed prior to the atomic model \cite{Drude-1900}. But, this kinetic model still treated individual electrons with classical properties. This model arrived at the important understanding that the effect of collisions was negligible. However, by not accounting for the quantum mechanical basis of the collisionless behavior, this model also erroneously predicted that the rate of collision increased with electron velocity.

Tomonaga \cite{Tomonaga-1950} first developed a quantum mechanical model of one-dimensional collective oscillations of Fermi electron gas. However, this model approximated the many-body wavefunction of the electron gas to be merely a collection of free electrons with a Fermi distribution at absolute zero, instead of electronic states with discrete wavevectors. In  \cite{Tomonaga-1950}, the degenerate electron gas is found to merely execute classical plasma oscillations in disagreement with experiments.

As is evident from models preceding Pines \cite{Pines-PhysRev-1953}, classical model contradict experiments while even simplified quantum models \cite{Pines-PhysRev-1953} are sufficient (including the effects of Bloch theorem, material dependence such as lattice structure etc.). The assumption that electrons in the presence of an ionic lattice merely form a classical fluid, entirely ignores the discrete electron states. 

Thus, classical approximations of a quantum system result in inaccurate predictions and are incapable of explaining experimental observations.

\subsection{Quantum coherence limit}
\label{subsec:quantum-coherence-limit}
Extreme plasmons \cite{plasmonics-ieee-2021,plasmonics-ieee-2022} for the first time access the coherence limit of  oscillations of the quantum electron gas that is inherent in conductive media. 

The conduction electrons that are excited to energy states, $\mathcal{E}\gg\mathcal{E}_F$, attain large $\vec{k_\ell}$. Collectively, electrons with various large $\vec{k_\ell}$ superpose to excite a large $\vec{k}$ plasmon, apparent from its field and density profile. 

For states with $\mathcal{E}\gg\mathcal{E}_F$, oscillations have large trajectories due to being strongly displaced from equilibrium. This increase in displacement requires a corresponding increase in restoring force exerted by the lattice to uphold the plasmon frequency, ${\rm \omega}_Q$,
\begin{align}\label{eq:plasmon-freq}
\begin{split}
	{\rm \omega}_{Q} & = \mathscr{F}_Q(\vec{k},\vec{p}) ~ \omega_{p} \\
							& \simeq \mathscr{F}_Q~ 56.4\sqrt{ n_0[10^{24}{\rm cm^{-3}}] } ~ {\times\rm 10^{15} rad/s} \\
\end{split}
\end{align}
While the quantum factor, $\mathscr{F}_Q$ depends on inherent characteristics of the material and properties of excitation, it primarily arises from the change in properties of a correlated many-body system relative to an ensemble of independent single electron states. It may be interpreted as the ability of quantum entities to be compressed to a density, $n_0=\mathscr{F}_Q^2n_e$ higher than uncorrelated ones. It is well known that a quantum ensemble becomes more collisionless at higher densities.

The restoring field that is capable of maintaining ${\rm \omega}_{Q}$ in Eq.\ref{eq:plasmon-freq} under coherence limited amplitude of oscillations of ultradense electron gas is,
\begin{align}\label{eq:plasmonic-field}
\begin{split}
	{\rm E}_Q & = \mathscr{F}_Q(\vec{k},\vec{p})\left(\frac{m_ec^2}{e}\right) \frac{2\pi}{\lambda_Q} \\
	 				& \simeq 0.1\mathscr{F}_Q\sqrt{ n_0[10^{24}{\rm cm^{-3}}] } ~ {\rm PVm^{-1}}.
\end{split}
\end{align}

The displaced Fermi electron gas that sustains the extreme plasmon, undergoes highly localized accumulation or compression which is ultimately limited by breaking of mutual coherence of plasmon under extreme amplitude, $\delta\simeq\lambda_Q$. This limit is a consequence of onset of trajectory overlap or collisions.

The density displaced $\delta n_e$ ($=n_e-n_0$) significantly exceeds equilibrium density, $\frac{\delta n_e}{n_0}\gtrsim 1$ owing to near complete displacement and compression as seen in Fig.\ref{fig:cross-section-3D-crunchin-plasmon}. Consequently, the ionic lattice gets locally bared by evacuation of the enveloping Fermi electron gas. The electric field in Eq.\ref{eq:plasmonic-field} is thus also the instantaneous Coulomb field of the unneutralized ionic lattice or the Fermi gas that get spatially segregated over plasmon timescales.

Eq.\ref{eq:plasmonic-field}, first introduced by our extreme plasmon model \cite{plasmonics-ieee-2021,plasmonics-ieee-2022, jinst-2023} is the electric field at the coherence limit of one-dimensional electron oscillations of a nonclassical gas. Although given the lack of appropriate experiments, Eq.\ref{eq:plasmonic-field} does not yet represent experimentally verified coherence limit of quantum electron oscillations. Specifically, a quantum gas where individual electrons are distinguishable with a well-defined quantum state. 

For a classical electron gas (Maxwell-Boltzmann statistics) commonly obtained by ionizing an atomic or molecular gas to plasma state, experiments have confirmed scaling of peak ${\rm E}_p$ \cite{wavebreaking-limit} with square root of ionized electron density, $\sqrt{n_e}$. Note that it is important to distinguish classical, ionized electron density, $n_e$ against quantum, Fermi electron density, $n_0$. Importantly, classical gasses can hardly be compressed to $n_e > 10^{19}{\rm cm^{-3}}$. 

In contrast, the quantum electron gas that is inherent in conductive materials is naturally ultradense ($n_0$) even at equilibrium. This is because the periodic ionic potential of the background lattice quantizes the electron energy levels in cognizance with the presence of lattice \cite{Bloch-1933}. The resulting electronic states are naturally independent of collisions with the lattice (without impurities, inconsistencies or thermal vibrations). 

Similarly, electron-electron collision cross-sections are suppressed due to a finite separation in electron energies especially owing to spin states, per the exclusion principle \cite{Pauli-spin-statistics}. Furthermore, correlation energy of electrons effectively allows higher spatial compression.

It is these implicit quantum effects that make possible existence of plasmons.

\subsection{Relativistic quantum dynamics} 

{\bf Relativistic quantum tunneling:} As the Fermi electron gas gains kinetic energy approaching (and eventually exceeding) the surface potential, the probability of its traversing the material interface through tunneling increases \cite{Fowler-Nordheim}. The rate of tunneling increases with magnitude of the applied electric field. In our model, electron tunneling which is an ultrafast process with attosecond timescale \cite{attosecond-tunneling}, occurs well within collisionless timescales. The conduction electron gas, thus, radially oscillates (see sec.\ref{sec:kinetic-model}) across the interface.

{\bf Relativistically induced ballistic transport:} It is well known that the mean free path, $\lambda_{\rm MFP}$ of conduction electrons is significantly higher than that of a classical electron gas only due to its quantum nature. Specifically, the Fermi-Dirac distribution sets up the average velocity around the Fermi velocity, $v_F\simeq0.01c$ at equilibrium, resulting in $\lambda_{\rm MFP}\gg d_0$.
 
Ultrafast excitation of the quantum gas from Fermi velocity, at equilibrium to relativistic velocities, correspondingly increases $\lambda_{\rm MFP}$ by about at least two orders of magnitude. This new effect of relativistically induced ballistic transport uncovered in our work is, thus a key characteristic of extreme plasmons. 

\section{Controlled excitation}
\label{sec:controlled-excitation}
Our work \cite{pct-2021} devises an experimentally realizable mechanism to controllably excite coherence-limited oscillations of quantum electron gas.

This is made possible using a new type of plasmon, the surface crunch-in plasmon \cite{ion-wake, crunch-in-regime}, which unlike conventional plasmons, allows sufficient control (under specific constraints per below) to make extreme plasmons and the PV/m field frontier realizable. 

The effectiveness of surface crunch-in plasmons lies in overcoming the constraints in experimental realization of extreme plasmons. Key requirements listed below not only allow practically viable nanometric confinement but also optimize energy exchange:
\begin{enumerate}[topsep=1pt, itemsep=0ex,partopsep=0ex,parsep=0ex]
\item {\em Collisionless} interaction such that the excitation propagates (in vacuum) along a conductive surface without  any direct contact with the material.
\item {\em Pre-pulse energy} (longer timescale) is minimized to avoid disruption of the properties of the material ahead of the main excitation.
\item {\em Maximize energy transfer} by enclosing the propagating excitation with quantum electron gas in surfaces that surround it.
\item {\em Matching} the properties of excitation with those of the material including its size, surface structure, conduction electron density ($n_0$) etc.
\item {\em Continual focusing} of a significant part of the excitation pulse to avoid dilution of its intensity owing to natural divergence.
\end{enumerate}

\noindent {\bf Collisionless excitation:} Direct interaction between the excitation pulse (charged particle or photon bunch) and the ionic lattice is highly disruptive. The pulse disintegrates due to rapid growth of numerous instabilities apart from uncontrolled, collisional loss of excitation energy. As a result, sustained interaction of charged particle beam with bulk solids has been impractical and experimentally unrealizable due to rapid breakup of the excitation\cite{filamentation}. A key requirement is, therefore, to mitigate direct collision.  

In our work, extreme plasmons are excited in a hollow tube such that the charged particle beam propagates inside the evacuated core. This substantially minimizes direct collision of the beam particles (except those in the wings) with the material in the wall.

\noindent {\bf Eliminate pre-pulse:} When pulses are temporally compressed to sub-picosecond timescales, their interaction with materials become quite distinct compared to nanosecond or longer pulses \cite{lasik-surgery}. 

However, ultrafast pulses are known to be accompanied by significant energy in pre-pulse ahead of the main pulse. This energy disrupts the material structure and its quantum nature. In a particle accelerator, an acceleration bucket which accepts a particle bunch inherently filter our any particles in the pre-pulse.
 
\noindent {\bf Maximize energy exchange:} In our work the quantum electron gas surrounds the excitation azimuthally enclosing it and participates in the collisionless energy exchange unlike a planar or bulk material.

\noindent {\bf Tunable nanomaterials:} Nanofabrication technology allows highly tunable properties of plasmonic materials. Structures with dimensions approaching the inherent scales of plasmons enable tunability of response of the media to external excitations.

\noindent {\bf Continual focusing:} Focusing the excitation pulse inside the tube is not possible using conventional surface modes or plasmons (such as the Transverse Magnetic mode). Moreover, these conventional ``purely'' electromagnetic modes of tubes are highly sensitive to transverse spatial or angular misalignments which disrupt and deflect the beam, a problem that rapidly grows with decreasing tube radius.

The surface ``crunch-in'' plasmon \cite{ion-wake,crunch-in-regime} addresses this critical need to continually focus the excitation (as well as a trailing bunch in the right phase). This becomes possible as the crunch-in plasmon is strongly electrostatic, unlike linear surface modes, and sustains strong focusing fields of the same order as Eq.\ref{eq:plasmonic-field}.

\section{Surface crunch-in plasmon}
\label{sec:kinetic-model}
 
In the analysis below we present the kinetic model of surface crunch-in plasmon excited by a charged particle beam. Our model is applicable over collisionless and coherent timescales of extreme plasmon. In consideration of the symmetry of interaction, with quantum electron gas in a tube surrounding the beam, our model is formulated in cylindrical coordinates.

{\bf Radial oscillations:} The charged particle beam which propagates in the positive $z$-direction is taken to be sufficiently relativistic, $\gamma_b\gg1$ such that its electric field, $E_b$ is primarily radial. Therefore, conduction electrons interacting with the radial fields of the beam predominantly oscillate radially. These radial collective oscillations of the quantum electron gas across the surface are significantly different from conventional plasmons. The beam density, $n_b$ ($E_b\propto n_b$) is sufficiently high to strongly excite conduction electrons, ${\rm max}[r(t)]-r_0\simeq\lambda_Q$ (see below) in less than, $2\pi\omega_Q^{-1}$.

{\bf Discrete momentum states:} While conduction electrons are delocalized and free to occupy any arbitrary radial position, $r(t)$, the same is not true about their momentum, $\vec{p}$ which is quantum mechanically, $\hbar\vec{k_\ell}$. The classical continuum in electron momentum, $\frac{dr(t)}{dt}$ is not valid in a quantum system. Therefore, a specific electron lattice wavevector, $\vec{k}_\ell$ needs to exist to accommodate an excited electron. 

However, to simplify our analysis we make a reasonable assumption that electrons excited such that they occupy large quantum numbers in the extreme limit, have discrete quantum states that can be approximated using a continuum. Under this approximation, below momentum states are simplified to be represented as $\frac{dr(t)}{dt}$. 

Nevertheless, the effect of ionic lattice on constraining occupancy to specific lattice wavevectors, $\vec{k}_\ell$, the effect of momentum, $\vec{p}$ and the effect of collective wavevector, $\vec{k}$ on quantum electron gas oscillations, is accounted for using the quantum factor, $\mathscr{F}_Q(\vec{p},\vec{k})$.

{\bf Mixed states:} The simplification afforded by assuming a continuum in momentum also helps in handling the challenge posed by conduction electrons tunneling across the surface. This changes (oscillates) their quantum nature (discrete states) in the presence of the lattice to being free electrons (continuum) in vacuum.

\begin{figure}[!htb]
\centering
   \includegraphics[width=0.75\columnwidth]{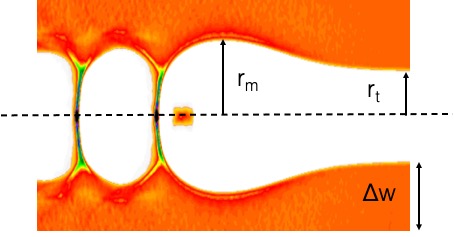}
   \caption{Notations in our surface crunch-in plasmon kinetic model. Here, $r_t$ is the tube radius, $r_m$ is the maximum radial amplitude of plasmon and $\Delta w$ is the wall thickness of the tube.This is zoomed-in version of Fig.\ref{fig:cross-section-3D-crunchin-plasmon}.}
\label{fig:3D-crunchin-mode-dimensions}
\end{figure}
The parameters of the geometry of conduction electron gas are (depicted in Fig.\ref{fig:3D-crunchin-mode-dimensions}):
\begin{enumerate}[label=\roman*. ,topsep=0pt, itemsep=0ex,partopsep=0ex,parsep=0ex]
\item $n_0$ is the conduction electron density.
\item $r_t$ is the radius of the vacuum core of the tube defined by its inner wall.
\item $\Delta w$ is the wall thickness with the corresponding outer wall radius being $r_t+\Delta w$.
\end{enumerate}

The excitation beam is assumed to approximate a Gaussian distribution (with azimuthal symmetry):
\begin{align}\label{eq:Gaussian-bunch}
\begin{split}
	n_b(r,z) = n_{b0} ~ {\rm exp}\left({-\frac{r^2}{2\sigma_r^2}}\right) ~ {\rm exp}\left({-\frac{(z-z_{\rm m})^2}{2\sigma_z^2}}\right)
	\end{split}
\end{align}
where the beam parameters are as follows:
\begin{enumerate}[label=\roman*. ,topsep=0pt, itemsep=0ex,partopsep=0ex,parsep=0ex]
\item $\sigma_z$ is the standard deviation along the longitudinal dimension of the Gaussian bunch and characterizes its bunch length.
\item $\sigma_r$ is the standard deviation along the radial dimension (under transverse symmetry) of the Gaussian bunch and characterizes the bunch waist-size at tube entrance.
\item $n_b(r,z)$ is the beam density ($n_{b0}=\frac{N_b}{(2\pi)^{3/2} \sigma_r^2 \sigma_z}$ is the peak density at $r=0$ and $z = z_{\rm m}$).
\item $N_b$ is the number of beam particles and $Q_b = {\rm sgn}[Q_b] ~ e ~ N_b$ is the bunch charge, and $N_b =  \int_{-\infty}^{\infty} \int_0^{\infty} \int_0^{2\pi}  n_b(r,z) ~ d\theta ~ r dr ~ dz$.
\item $\vec{v_b}=\boldsymbol\beta_b c$ is beam velocity and $\gamma_b = (1-\beta_b^2)^{-1/2}$.
\end{enumerate}

The electric field lines of a gaussian bunch of ultra-relativistic charged particles, $\gamma_b\gg1$ are predominantly radial and perpendicular to the direction of the bunch velocity, $|\boldsymbol\beta_b|=\sqrt{1 - 1/\gamma_b^2}$. The radial electric field $E_b(r, z)$ of the beam assuming azimuthal symmetry and $\gamma_b\gg1$ is obtained using Gauss's law,
\begin{align}\label{eq:fields-Gaussian-bunch}
\begin{split}
E_{\rm b}&(r, z) = - {\rm sgn}[Q_b] ~ \omega_{p}^2(n_0) ~ \frac{m_e}{e}\frac{n_{b0}}{n_0} ~ \times \\
& \qquad \frac{\sigma_r^2}{2\pi} ~ \frac{\left[ 1 - {\rm exp}\left( -\frac{r^2}{2\sigma_r^2} \right) \right] }{r} ~ {\rm exp}\left({-\frac{(z-z_{\rm m})^2}{2\sigma_z^2}}\right)
	\end{split}
\end{align}
The beam is here assumed to be in a classical state (plasma, $\omega_{p}$) but it may also attain a quantum state when it satisfies the conditions on $\Gamma$ and $\chi$ in sec.\ref{subsec:quantum-state}.

As the beam undergoes focusing being acted upon by the radial fields of the extreme plasmon, its profile gets modified and the field changes to $- \mathrm{ sgn}[Q_b] ~ \omega_{p}^2(n_0) ~ \frac{m_e}{e} ~ \frac{n_{b0}(\xi)}{n_0} ~ \int_{0}^{r} dr ~ \mathcal{P}(r,z,\xi)$, where $\mathcal{P}(r,z,\xi)$ is the profile function of the continually focused beam. Here, we simplify the analysis assuming that the profile changes slowly and remains Gaussian.

\subsection{Quantum kinetic equation using $\mathscr{F}_Q(\vec{k},\vec{p})$}

A kinetic equation is developed here within the quantum kinetic framework using modified single-particle approach (sec.\ref{quantum-kinetic-approx}). Specifically, an equation of motion of individual conduction electrons oscillating at average frequency, $\omega_Q$ is obtained. 

Assuming azimuthal symmetry, radial motion over an infinitesimally thin disk of thickness $dz$ is considered. The key parameters of the kinetic model are:
\begin{description}[itemsep=0pt, topsep=0pt, parsep=0pt]
\item[$r_0$] is the equilibrium position of a tube electron which always satisfies the condition, $r_t < r_0 < r_t+\Delta w$
\item[$r(t)$] is the instantaneous radial position of an oscillating tube electron ($r = r_0 + \Delta r$ where, $\Delta r$ is the radial displacement of the electron which can be positive or negative)
\item[$r_m$] is the maximum radius to which the strongly driven tube electrons reach and bunch up to form an electron sheath 
\item[$\mathcal{H}(r)$] is the step function which is defined as $\mathcal H(0^+) = 1$ and $\mathcal H(0^-) = 0$. This function models the step-like change in the density across the surface.
\end{description}

{\bf Coherent oscillations:}  Radially oscillating conduction electrons retain mutual coherence until they maintain the same ordering in the radial direction, as that at equilibrium. Ordering precludes crossing of trajectories and disallows electron-electron collisions, a key requirement for ballistic oscillations.

This condition requires that a conduction electron located at an equilibrium radial position, $r_0$ which is initially less than that of a neighboring electron, always remains at a lesser instantaneous radii, $r(t)$ throughout its trajectory underlying the plasmon. 

It requires that when other conduction electrons which are located at an equilibrium radial position less than the electron under consideration (between $r_0$ and $r_t$) move with it, they remain ordered and just behind $r(t)$. This adherence to equilibrium ordering during oscillation ensures coherent and collisionless oscillations.

Mutual coherence along with collisionless characteristics ensure that the plasmon does not collisionally dissipate energy coupled to it.

{\bf Ionic lattice phase:} When a conduction electron under consideration is radially displaced from $r_0$ (by the force of beam fields), it uncovers the ionic lattice in the tube wall. Upon radial displacement of this electron, all the electrons located at equilibrium radii between $r_t$ and $r_0$ move together and are displaced to a new radial location $r(t)$. These electrons bunch together into a sheath layer at the trajectory maxima due to inward force of the uncovered ionic lattice. The inward force exerted due to uncovering of the ionic lattice (using Gauss's law) is:
\begin{align}\label{eq:ionic-force-r}
\begin{split}
-e ~ E_{\rm lattice}(r > r_t) = - m_e \omega_{p}^2(n_0) ~ \frac{(r^2 - r_t^2)}{2r}
\end{split}
\end{align}

However, in addition to the ionic force on the electron under consideration (originally located at $r_0$), the electrons with an equilibrium radii smaller than $r_0$ (located between $r_0$ and $r_t$) collectively move with the electron under consideration and compress together to produce a collective field opposite to that of the lattice. The compressed density is higher in the quantum regime by $\mathscr{F}_Q^2$. The force of electrons collectively compressed together, balances the force exerted by the lattice to form dynamic equilibria (zero velocity) at the trajectory maxima, $r_m$. 

The field due to accumulation of conduction electrons located between $r_0$ and $r_t$ and collectively moving outwards along with the electron at $r_0$ while maintaining the equilibrium radial ordering is,
\begin{align}\label{eq:electronic-force-r}
\begin{split}
-e ~ E_{\rm e}(r) = m_e\omega_{Q}^2(n_0) ~ \frac{(r_0^2 - r_t^2)}{2r}
\end{split}
\end{align}
As conduction electrons undergo a compression greater by $\mathscr{F}_Q^2n_0$, the field is correspondingly higher.

{\bf Crunch-in phase:} After the equilibrium at the outward radial maxima, $r_m$ electron momentum becomes radially negative. Being part of large-amplitude oscillations, the electron under consideration gains large enough velocity to tunnel across the surface and move into the hollow region of the tube. As there is no lattice in the core region of the tube, the collectively moving electrons do not experience any ionic force.

However, since the electrons collectively oscillate, the tube electrons located between $r_0$ and $r_t$ collectively crunch-in into the hollow region. The stronger the force of the beam, the deeper is the radial excursion into the core region of the tube. Under this condition, inside the tube core region the only force acting to restore the electrons back to equilibrium is due to electron accumulation. This is equivalent to the force in Eq.\ref{eq:electronic-force-r}.

Therefore, the net collective force acting on the electrons can be written using the Heaviside step function, $\mathcal{H}(r)$ to segregate the forces controlling the dynamics in the wall in contrast with that inside the tube:
\begin{align}\label{eq:net-force-r}
\begin{split}
F_{\rm coll} &(r>r_t) = -e ~ E_{\rm lattice}(r > r_t) -e ~ E_{\rm e}(r > r_t) \\
								& = - m_e \omega_{Q}^2(n_0) \frac{ (r^2 - r_t^2) \mathcal{H}(r-r_t) - (r_0^2 - r_t^2)  }{2r}
\end{split}
\end{align}
$\mathcal{H}(r-r_t)$ in Eq.\ref{eq:net-force-r} indicates that the first term which is due to the lattice along with the second term together act on the electrons while their instantaneous radial position is in the wall. However, the second term only applies to electrons with instantaneous position in the evacuated region, $r < r_t$, as the effect of lattice ceases at $r_t$.

It is important to note that due to the termination of the lattice at $r=r_t$, the kinetic equation does not account for quantum effects inside the tube core. But, continuity of momentum across the boundary needs to be taken into account. An additional term, $\frac{\partial}{\partial t}\left(\frac{\partial r}{\partial t}\right) |_{r=r_t}$ is used to take this continuity in momentum.

{\bf Beam force:} When negatively charged, ${\rm sgn}[Q_b] = -1$ beam, such as an {\it electron beam}, excites the conduction electrons, they initially propagate outwards and compress together at $r_m$. On the other hand, when excited by a positively charged beam, conduction electrons initially accelerate inwards towards the core region. The bunch length is taken to be smaller than, $\lambda_Q$ such that the beam fields terminate in the crunch-in phase. 
 
{\bf Kinetic equation:} For a conduction electron located at $r(z,t)$, the equation of motion of collective surface oscillation excited by the beam field is $\gamma_e m_e \frac{\partial^2 r(z,t)}{\partial t^2} = F_{\rm coll} + eE_b$. Here, $\gamma_e = \left( 1 + \frac{\vec{p}}{m_ec}\cdot\frac{\vec{p}}{m_ec}\right)^{1/2}$ is the relativistic factor of the conduction electrons and beam field is simplified under the condition, $\sigma_r\lesssim r_t$. 

The kinetic equation of the nonlinear surface wave sustained by localized crunch-in plasmons is obtained by transforming to a coordinate co-moving with the beam, $\xi = c\beta_b t - z$ such that $\frac{\partial \xi}{c\beta_b} =  \partial t$ and using Eq.\ref{eq:fields-Gaussian-bunch},
\begin{align}\label{eq:crunch-in-plasmon-kinetic}
\begin{split}
\frac{\partial^2 r(z,t)}{\partial \xi^2} & +  \frac{\omega_{p}^2(n_t)}{\gamma_e c^2\beta_b^2} \frac{1}{2r} \left[ (r^2 - r_t^2)\mathcal{H}(r-r_t) - \mathscr{F}_Q^2(r_0^2 - r_t^2)  \right] \\
& + {\rm sgn} \left[\frac{\partial r}{\partial \xi}\right] ~ \frac{\partial^2 r}{\partial \xi^2} \biggm\lvert_{r=r_t}  \\
& = - {\rm sgn}[Q_b] \frac{\omega_{p}^2(n_0)}{\gamma_ec^2\beta_b^2} ~ \frac{n_{b0}}{n_0} \frac{\sigma_r^2}{2\pi} ~ \frac{1}{r} ~ e^{{-(z-z_{\rm m})^2/(2\sigma_z^2)}} 
\end{split}
\end{align}

The kinetic equation, Eq.\ref{eq:crunch-in-plasmon-kinetic} is used obtain the radial maxima of electron trajectory, $r = r_m$ where $\frac{\partial^2 r}{\partial t^2} = 0$,
\begin{align}\label{eq:r-max-nano-Gaussian}
\begin{split} 
r_m ~ = ~\left[ r_t^2 ~ + 2 ~ \frac{n_{b0}}{n_0} ~ \frac{\sigma_r^2}{2\pi} \right]^{1/2} 
\end{split}
\end{align}

Existence condition of crunch-in plasmon,  $(r_t+\Delta w)>r_m$ is also obtained as,
\begin{align}\label{eq:existence-condition}
\begin{split} 
\Delta w \left(1+\frac{\Delta w}{2r_t}\right) > \frac{n_0}{n_{b0}} ~ \frac{\sigma_r^2}{2\pi r_t}
\end{split}
\end{align}

Extreme plasmon is excited only when $(r_m - r_t)\lambda_Q^{-1}\gg 0$, which is simplified as,
\begin{align}\label{eq:excitation-condition}
\begin{split}
0\ll\left( \frac{n_{b0}}{n_0} ~ \frac{\sigma_r^2}{2\pi r_t} \right) \lambda_Q^{-1} \simeq 1
 \end{split}
\end{align}

\section{Fields of the extreme plasmon}
\label{sec:field-estimation}

When the existence, Eq.\ref{eq:existence-condition} and excitation condition, Eq.\ref{eq:excitation-condition} are satisfied, the kinetic model in Eq.\ref{eq:crunch-in-plasmon-kinetic} and the expression for $r_m$ in Eq.\ref{eq:r-max-nano-Gaussian}, allow determination of the net charge displaced in the plasmon. Using the net charge that crunches-in into the core region of the tube, magnitude of peak fields is obtained.

The total charge that builds up in the electron compression layer at $r_m$ during the outward radial excursion phase, is restored back towards the axis by the force of the uncovered ionic lattice. However, instead of merely reverting to their respective equilibrium positions, the excited electrons collectively cross the surface to crunch in towards the axis, $r=0$. 

During this crunch-in phase of radially inward electron compression, the force of ionic lattice ceases to exist for $r < r_t$. So, electrons continue to crunch in until the total compressed electron density results in a radial electric field, $E_r$ that opposes further compression. The electrons first crunch in to a minimum radius, $r_{c}$ at the longitudinal position, $\xi_{r_c}$. This minimum radius, $r_{c}$ is dictated by their initial acceleration at excitation.

The charge density that crunches in within the radial area enclosed by $r_{c}$ is calculated using the collective motion of all electrons between $r_m$ and $r_t$ to result in their collective collapse into the tube. 

The net change displaced in the excited electron rings with infinitesimal slice thickness, $dz$, that crunch-in at the longitudinal position, $\xi_{r_c}$ of maximum inward compression (equal to the charge at maximum outward compression) using Eq.\ref{eq:r-max-nano-Gaussian} is,
\begin{align}\label{eq:charge-collective-rings-step-nano-Gaussian}
\begin{split}
\delta Q_{\rm m}&(\xi_{r_c})  = - e n_0 \pi (r_m^2 - r_t^2) ~ dz = - e n_{b0} \sigma_r^2 ~ dz
\end{split}
\end{align}

The minimum radius to which the electrons crunch-in to is represented relative to the tube radius, $r_{c} = r_t / (\mathscr{F}_Q\Theta)$, where $\Theta>1$. The minimum radius to which the electrons compress increases by $\mathscr{F}_Q$. The peak radial electric field using the net charge, $Q_m$, and its simplification using SI units is,
\begin{align}\label{eq:crunch-in-charge-radial-field-Gaussian}
\begin{split}
E_{r}(\xi_{r_c}) & = -\mathscr{F}_Q ~ \frac{2\Theta}{(2\pi)^{3/2}} ~ \frac{1}{r_t} ~ \frac{Q_b}{\sigma_z} \\
& = - \mathscr{F}_Q ~ \Theta ~ \frac{114.2}{r_t\rm[100nm]}  ~ \frac{Q_b{\rm [nC]} }{\sigma_z{\rm [100nm]} } {\rm \frac{TV}{m}}
\end{split}
\end{align}
Because the beam field, $E_b(r)$ outside the beam decreases radially, conduction electrons at increasing radial equilibrium positions $r_0>r_t$, gain decreasing initial momentum. Thereby, when electrons crunch into the tube each reaches a different crunch-in radius, $r_c(r_0,t)$ at any instant. $E_r(r)$, therefore, increases radially outwards until $r_t$, as the net crunched in charge increases. Consequently, $\Theta$ depends upon the excitation conditions.

The peak longitudinal field, $E_{z}$ is obtained using the Panofsky-Wenzel theorem \cite{Panofsky-Wenzel-theorem}, $\frac{1}{r}\frac{\partial rE_z}{\partial r} = -\frac{\partial E_r}{\partial \xi}$ or ${\rm lim_{\Delta\rightarrow 0}}\frac{\Delta r E_z}{\Delta r} = -r\frac{\Delta E_r}{\Delta \xi}$. $E_{z}$ radially varies over $\Delta r\simeq r_t$ as the phase of the radial field transitions from negative to positive over $\Delta \xi=\kappa \lambda_c$. The crunch-in wavelength for a given set of beam properties, $\lambda_c$ is itself proportional to the tube radius $r_t$ \cite{plasmonics-ieee-2022}. As the tube radius increases, per the excitation condition in Eq.\ref{eq:excitation-condition} the plasmon becomes less nonlinear. On the other hand, as $r_t\rightarrow 0$ (bulk media) $\lambda_c\simeq\lambda_Q$. Therefore, relativistically corrected wavelength of the surface crunch-in plasmon is, $\lambda_c = \lambda_Q \sqrt{\langle\gamma_e\rangle} ~ \left[ \frac{r_t}{\lambda_Q} + 1\right]$. 

An approximate expression for the average $\langle\gamma_e\rangle$ is obtained using $\Delta p_r = eE_b\frac{\sigma_z}{c} \gtrsim m_ec$, such that,
$\gamma_e = \sqrt{1 + \left(\frac{p_r}{m_ec}\right)^2 } \simeq \frac{\omega_{p}^2(n_0)}{c^2} ~ \frac{1}{r_t} ~ \frac{n_{b0}}{n_0} ~ \frac{\sigma_r^2\sigma_z}{2\pi}$. 
Using this $\gamma_e$, the peak longitudinal field, $E_{z}$ is evaluated as:
\begin{align}\label{eq:crunch-in-panofsky-wenzel-simplified-sigr-lte-rt}
\begin{split}
E_z & = E_{r} ~ \frac{r_t}{\kappa ~ 2\pi c} \frac{\omega_{Q}(n_0)}{\sqrt{\gamma_e} }
= E_{r} ~ \frac{r_{t}}{\kappa \sqrt{2\pi}} \sqrt{ \frac{n_0 r_t}{n_{b0} \sigma_r^2\sigma_z } }
\end{split}
\end{align}
The peak longitudinal field, $E_{z}(\xi_{r_c})$ is obtained using Eq.\ref{eq:crunch-in-charge-radial-field-Gaussian} with $\left(1+r_t/\lambda_Q\right)\simeq 1$ and simplified using SI units,
\begin{align}\label{eq:crunch-in-charge-acc-field-Gaussian}
\begin{split}
E_{z}&(\xi_{r_c}) 
= \frac{\mathscr{F}_Q}{\kappa} ~ \frac{\sqrt{2}}{(2\pi)^{7/4}} ~ \sqrt{\frac{Q_b}{e}} ~ \frac{\sqrt{r_t r_e}}{\sigma_z} ~ E_{Q} \\
& = \mathscr{F}_Q^2 ~ \frac{3}{4\kappa} \sqrt{n_0[{\rm cm^{-3}}] ~ r_t\rm[100nm]} \frac{\sqrt{Q_b{\rm [nC]}} }{\sigma_z{\rm [100nm]} } {\rm \frac{TV}{m}}
\end{split}
\end{align}

The expressions for peak electric fields in Eq.\ref{eq:crunch-in-charge-radial-field-Gaussian} and Eq.\ref{eq:crunch-in-charge-acc-field-Gaussian} are evaluated against previously published simulation results based on classical kinetic theory. 

From the nanoporous metal simulations in \cite{plasmonics-ieee-2021}, using $n_0 = 2\times 10^{22} {\rm cm^{-3}}$, $r_t = 100 {\rm nm}$, $Q_b = 315 ~ {\rm pC}$, and $\sigma_z=400{\rm nm}$; the peak radial field is, $E_r = \mathscr{F}_Q ~ \Theta ~ 9 ~ {\rm TV/m}$ and longitudinal field with $\kappa=0.25$ is, $E_z = \frac{\mathscr{F}_Q}{\kappa} 5.74 ~ {\rm TV/m}$. Similarly, for semiconductor plasmon simulations in \cite{plasmonics-ieee-2022}, with $n_0 = \mathrm{10^{18} cm^{-3}}$, $r_t = \mathrm{12.5\mu m}$, $n_{b0} = \mathrm{10^{18} cm^{-3}}$, $Q_b = 1.5 ~ {\rm nC}$ and $\sigma_z \simeq \mathrm{10 \mu m}$; the peak radial field is, $E_r = \mathscr{F}_Q ~ \Theta ~ 13.7 ~ {\rm GV/m}$ and peak longitudinal field with $\kappa=0.25$ is, $E_z = \frac{\mathscr{F}_Q}{\kappa} 39.4 ~ {\rm GV/m}$. This establishes a good agreement with both previously published simulation results validating the quantum model of extreme plasmon presented here. 

However, precise characteristics including $\Theta$, $\kappa$, and especially $\mathscr{F}_Q$ that emanates from the quantum nature need dedicated efforts to develop a comprehensive understanding. Our ongoing work including experiments \cite{slac-2023,slac-2022,slac-2020} is geared towards addressing this.

\section{Conclusion}

In conclusion, characteristics of the new class of plasmons necessitate a non-perturbative, quantum kinetic framework. 

A modified independent electron approximation introduced here that accounts for quantum frequency of plasmons, $\omega_Q$ using, $\mathscr{F}_Q(\vec{k},\vec{p})$, is found to be valuable to simplify the underlying non-perturbative, collisionless and relativistic quantum dynamics. The kinetic model of experimentally realizable extreme plasmon, the surface crunch-in plasmon, based on this approximation provides the existence and excitation conditions, in addition to expressions of its focusing and longitudinal fields.

For further strengthening the extreme plasmon model towards unprecedented $\rm PVm^{-1}$ fields, desired experimental efforts are ongoing \cite{slac-2023,slac-2022,slac-2020}.

\acknowledgments
This work was supported by the department of Electrical Engineering at University of Colorado Denver. The use of open-source particle tracking code EPOCH \cite{epoch-pic} is acknowledged. The RMACC Summit and Alpine supercomputers utilized in this work were supported by the NSF awards ACI-1548562, ACI-1532235 and ACI-1532236, the University of Colorado Boulder, and Colorado State University \cite{xsede-rmacc-citation}.



\begin{thebibliography}{99}

\bibitem{Pines-Bohm-PhysRev-1952}
Bohm, D., Pines, D.,
	\textit{A Collective Description of Electron Interactions: III. Coulomb Interactions in a Degenerate Electron Gas},
	\href{https://doi.org/10.1103/PhysRev.92.609}{Phys. Rev. {\bf 92}, 609 (1953)}
	
\bibitem{Pines-PhysRev-1953}
Pines, D.,
	\textit{A Collective Description of Electron Interactions: IV. Electron Interaction in Metals},
	\href{https://doi.org/10.1103/PhysRev.92.626}{Phys. Rev. {\bf 92}, 626 (1953)}
	
\bibitem{Ritchie-surface-plasmon}
Ritchie, R. H.,
	\textit{Plasma Losses by Fast Electrons in Thin Films},
	\href{https://doi.org/10.1103/PhysRev.106.874}{Phys. Rev. {\bf 106}, 874 (1957)}

\bibitem{plasmonics-1}
Stockman, M., Kneipp, K., Bozhevolnyi, S., Saha, S., Dutta, A., et. al.,
	\textit{Roadmap on plasmonics},
	\href{https://doi.org/10.1088/2040-8986/aaa114}{J. Opt. {\bf 20} 043001 (2018)}
	
\bibitem{Bloch-1933}
Bloch, F.,
	\textit{Bremsvermogen von Atomen mit mehreren Elektronen},
	\href{https://doi.org/10.1007/BF01344553}{Zeitschrift fur Physik, {\bf 81}, iss. 5-6, pp. 363-376 (1933)}
	
\bibitem{Pauli-spin-statistics}
Pauli, W.,
	\textit{The Connection Between Spin and Statistics},
	\href{https://doi.org/10.1103/PhysRev.58.716}{Phys. Rev. {\bf 58}, 716 (1940)}

	
\bibitem{plasmonics-ieee-2021} Sahai, A. A., 
	\textit{Nanomaterials Based Nanoplasmonic Accelerators and Light-Sources Driven by Particle-Beams}, 
	\href{https://doi.org/10.1109/ACCESS.2021.3070798}{IEEE Access, {\bf 9}, pp. 54831-54839 (2021), doi: 10.1109/ACCESS.2021.3070798}
	
\bibitem{spie-2021} Sahai, A. A., 
	\textit{Emergence of TeraVolts per meter plasmonics using relativistic surface plasmons},
	\href{https://doi.org/10.1117/12.2596637}{Proc. Vol. {\bf 11797}, Plasmonics: Design, Materials, Fabrication, Characterization, and Applications XIX; 117972A (2021)}

\bibitem{pct-2021} Sahai, A. A.,
	\textit{Nanostructure nanoplasmonic accelerator, high-energy photon source, and related methods},
	\href{https://patentimages.storage.googleapis.com/55/0a/6c/7b757efb9e9547/WO2021216424A1.pdf}{PCT WO2021216424A1}
	
\bibitem{plasmonics-ieee-2022} Sahai, A. A., Golkowski, M., Katsouleas, T., 
	\textit{Approaching PetaVolts per Meter Plasmonics Using Structured Semiconductors},
	 \href{https://doi.org/10.1109/ACCESS.2022.3231481}{IEEE Access, {\bf 11}, pp. 476-493, (2023), doi: 10.1109/ACCESS.2022.3231481}
	 
\bibitem{ion-wake}
Sahai, A. A.,
	\textit{Excitation of a nonlinear plasma ion wake by intense energy sources with applications to the crunch-in regime},
	\href{https://doi.org/10.1103/PhysRevAccelBeams.20.081004}{Phys. Rev. Accel. Beams {\bf 20}, 081004 (2017)}

\bibitem{crunch-in-regime}
Sahai, A. A.
	 \textit{Chapter 8, On Certain Non-linear and Relativistic Effects in Plasma-based Particle Acceleration},
	 \href{hhttps://dukespace.lib.duke.edu/dspace/handle/10161/10534}{Chap. 8, Ph.D. thesis, Duke University, 2015, 3719664}
	
\bibitem{nanofocusing-2022}
Sahai, A. A.,
	\textit{Plasmonic nano-focusing of particle beams by surface crunch-in plasmons excited in tapered tubes},
	\href{https://doi.org/10.1117/12.2605722}{Proc. SPIE {\bf 11999}, Ultrafast Phenomena and Nanophotonics XXVI, 1199903 (2022)}
	
\bibitem{jinst-2023}
Sahai, A. A., Golkowski, A. A., et. al.,
	\textit{PetaVolts per meter Plasmonics: introducing extreme nanoscience as a route towards scientific frontiers},
	\href{https://doi.org/10.1088/1748-0221/18/07/P07019}{Journal of Instrumentation {\bf 18}, P07019 (2023)}
	
\bibitem{ginzburg-2003} Ginzburg, V. L.,
	\textit{Nobel lecture: On superconductivity and superfluidity},
	\href{https://doi.org/10.1103/RevModPhys.76.981}{Reviews of Modern Physics, {\bf 75}, no. 3, p. 996, (2004)}	
	
\bibitem{schwinger-limit}
Schwinger, J.,
	\textit{On Gauge Invariance and Vacuum Polarization},
	\href{https://doi.org/10.1103/PhysRev.82.664}{Phys. Rev. {\bf 82}, 664 (1951)}
	
\bibitem{slac-2023}
Sahai, A. A.,
	\textit{PetaVolts per meter plasmonics},
	\href{https://facet-ii.slac.stanford.edu/sites/default/files/2023-10/PetaVolts per meter Plasmonics_SAHAI.pdf}{FACET-II science meeting, Oct. 2023}
	
\bibitem{slac-2022}
Sahai, A. A.,
	\textit{PetaVolts per meter plasmonics using structured semiconductors},
	\href{https://www-lcls.slac.stanford.edu/web/FACET/pac meetings/pac2022/SAHAI_Plasmonics_FACET-II_2022_Nano2WA_proposal_0.pdf}{FACET-II user meeting, Oct. 2022}

\bibitem{slac-2020}
Sahai, A. A.,
	\textit{Ultra-solid beams using nanostructure nanoplasmonic wiggler and accelerator},
	\href{https://www-lcls.slac.stanford.edu/web/FACET/pac meetings/pac2020/FACET-II_PAC_committee_answers_nano2WA_proposal_Sahai.pdf}{FACET-II user meeting, Oct. 2020}

\bibitem{Wigner}
Wigner, E.,
	\textit{On the Quantum Correction For Thermodynamic Equilibrium},
	\href{https://doi.org/10.1103/PhysRev.40.749}{Phys. Rev. {\bf 40}, 749 (1932)}
	
\bibitem{Kronig-Korringa-1943}
Kronig, von R., Korringa, J.,
	\textit{Zur Theorie der Bremsung Schneller Geladener Teilchen in Metallischen Leitern},
	\href{https://doi.org/10.1016/S0031-8914(43)90032-1}{Physica {\bf 10}, iss. 6, pp.406-418 (1943)}
	
\bibitem{Drude-1900}
Drude, P., 
	\textit{Zur Elektronentheorie der Metalle},
	\href{https://doi.org/10.1002/andp.19003060312}{Ann. Phys., {\bf 306}, pp.566-613 (1900)}

\bibitem{Kramers-1947}
Kramers, H.A.,
	\textit{The stopping power of a metal for alpha-particles},
	\href{https://doi.org/10.1016/0031-8914(47)90014-1}{Physica {\bf 13}, iss. 6-7, pp.401-412 (1947)}
	
\bibitem{Tomonaga-1950}
Tomonaga, S.,
	\textit{Remarks on Bloch's Method of Sound Waves applied to Many-Fermion Problems},
	\href{https://doi.org/10.1143/ptp/5.4.544}{Progress of Theoretical Physics {\bf 5}, iss. 4, pp.544-569 (1950)}
		
\bibitem{wavebreaking-limit}
Dawson, J. M.,
	\textit{Nonlinear Electron Oscillations in a Cold Plasma},
	\href{https://doi.org/10.1103/PhysRev.113.383}{Phys. Rev. {\bf 113}, 383 (1959)}
	
\bibitem{filamentation}	
Benedetti, A., Tamburini, M., Keitel, C. H.,
	\textit{Giant collimated gamma-ray flashes},
	\href{https://doi.org/10.1038/s41566-018-0139-y}{Nature Photonics {\bf 12}, pp.319-323 (2018)}
	 
\bibitem{lasik-surgery}
Asbury, M.,
	\textit{How a Lab Incident Led to Better Eye Surgery for Millions of People},
	\href{https://www.goldengooseaward.org/01awardees/lasik}{American Association for the Advancement of Science, 2022}
	
\bibitem{Fowler-Nordheim}
Fowler, R. H., Nordheim, L. W.,
	\textit{Electron emission in intense electric fields},
	\href{https://doi.org/10.1098/rspa.1928.0091}{Proc. Roy. Soc. Lond. A {\bf 119}, p.173-181 (1928)}
	
\bibitem{attosecond-tunneling}
Sainadh, U.S., Xu, H., Wang, X. et al. 
	\textit{Attosecond angular streaking and tunnelling time in atomic hydrogen},
	\href{https://doi.org/10.1038/s41586-019-1028-3}{Nature {\bf 568}, 75-77 (2019)}.
	
\bibitem{Panofsky-Wenzel-theorem}
Panofsky, W. K. H., Wenzel, W. A.,
	\textit{Some Considerations Concerning the Transverse Deflection of Charged Particles in Radio-Frequency Fields},
	\href{https://doi.org/10.1063/1.1715427}{Review of Scientific Instruments {\bf 27}, 967 (1956)}
	
\bibitem{epoch-pic}
Arber, T. D., Bennett, K., K. Brady, K., Lawrence-Douglas,  A., Ramsay, M. G., et. al.,
	\textit{Contemporary particle-in-cell approach to laser-plasma modelling},
	\href{https://doi.org/10.1088/0741-3335/57/11/113001}{Plasma Phys. Control. Fusion, {\bf 57}, 113001 (2015)}
	
\bibitem{xsede-rmacc-citation}
Towns, J., Cockerill, T., Dahan, M., Foster, I., Gaither, K., et. al.,
	\textit{XSEDE: Accelerating Scientific Discovery},
	\href{https://doi.org/10.1109/MCSE.2014.80}{Computing in Science \& Engg. {\bf 16}, pp. 62-74, (2014)}
	
\end{thebibliography}
\end{document}